\documentstyle[12pt,a4]{article}

\newcommand{\be}{\begin{equation}} \newcommand{\ee}{\end{equation}}
\newcommand{\ba}{\begin{eqnarray}} \newcommand{\ea}{\end{eqnarray}}

\newcommand{\no}{\nonumber}
\newcommand{\Eqn}[1]{&\hspace{-0.5em}#1\hspace{-0.5em}&}

\newcommand{\WOaEei}[9]{{}\widehat{w}^{(#1)}
                          [{}_{#2}^{}{}_{#4}^{}{}_{#5}^{#3}{}_{#6}^{}
                           {}_{#7}^{}{}_{#8}^{}{}_{#9}^{}]}
\newcommand{\WOEei}[8]{{}w[{}_{#1}^{}{}_{#3}^{}{}_{#4}^{#2}{}_{#5}^{}
                           {}_{#6}^{}{}_{#7}^{}{}_{#8}^{}]}
\newcommand{\WOEse}[7]{{}w[{}_{#1}^{}{}_{#3}^{}{}_{#4}^{#2}{}_{#5}^{}
                           {}_{#6}^{}{}_{#7}^{}]}
\newcommand{\WOEsi}[6]{{}w[{}_{#1}^{}{}_{#3}^{}{}_{#4}^{#2}{}_{#5}^{}
                           {}_{#6}^{}]}
\newcommand{\tfrac}[2]{{\textstyle\frac{#1}{#2}}}
\newcommand{\tchara}[1]{\mbox{\footnotesize $#1$}}

\makeatletter
\@addtoreset{equation}{subsection}

\makeatother
\makeatletter
\@addtoreset{subsection}{section}
\def\thesubsection{\arabic{subsection}.}
\makeatother
\begin{document}

\vskip 7mm
%%% Title page %%%%%
\begin{titlepage}
 
 \renewcommand{\thefootnote}{\fnsymbol{footnote}}
 \font\csc=cmcsc10 scaled\magstep1
 {\baselineskip=14pt
 \rightline{
 \vbox{\hbox{hep-th/0203025}
       \hbox{UT-975}
       }}}

% \vfill
 \vskip4cm 
\baselineskip=20pt
 \begin{center}
 \font\titlerm=cmr10 scaled\magstep4
 \font\titlei=cmmi10 scaled\magstep4
 \centerline{\titlerm Seiberg--Witten Curve for the
                      $\mbox{\titlei E}$-String Theory} 
 \vskip 2.5 truecm

{\it \large Tohru Eguchi and Kazuhiro Sakai}
\bigskip

 \vskip .6 truecm
 {
 {\it Department of Physics,  Faculty of Science,\\
  University of Tokyo\\
  Tokyo 113-0033, Japan}
 }
 \vskip .4 truecm
%E-mail: {\tt eguchi,sakai@hep-th.phys.s.u-tokyo.ac.jp}

 \end{center}

 \vskip 2.5 truecm

\begin{abstract}

We construct the Seiberg--Witten curve for the $E$-string theory in
six-dimensions.
The curve is expressed in terms of affine $E_8$ characters up to level 6
and is determined by using the mirror-type transformation so that it
reproduces the number of holomorphic curves in the Calabi--Yau manifold
and the amplitudes of ${\cal N}=4$ $U(n)$ Yang--Mills theory
on ${1\over 2}K_3$. We also show that our curve flows to known five- and
four-dimensional Seiberg--Witten curves in suitable limits.

\end{abstract}

\end{titlepage}

\subsection{Introduction}

$E$-string was first discovered in the study of zero-size instantons in heterotic
string theory \cite{GH}. In the M-theory description \cite{HW} 
$E$-string appears at the intersection of the 
(end-of-world) 9-brane and a membrane which connects the 9-brane to a 5-brane 
representing 
the small instanton. An 
$E$-string carries a level one $E_8$ current algebra and possesses 
one half of the degrees of freedom of the heterotic string. 
In fact an $E$-string may be obtained by wrapping a M5-brane around ${1\over 2}K_3$ 
surface \cite{MNVW} while a heterotic string is obtained by wrapping 
M5-brane around $K_3$. The second homology $b_2^+=1,b_2^-=9$ of ${1\over 2}K_3$ 
generates 1 (9) right-(left-)moving bosonic degrees of freedom in 2-dimensions.
Together with the freedom representing the position of the M5-brane one obtains
4 (12) right (left) bosonic degrees of freedom of $E$-string which is one 
half of the
heterotic string. $E$-string possesses one half of supersymmetry of the
heterotic theory and 
the reduction of supersymmetry leads to threshold BPS bound states and  
non-trivial dynamics in low energy $E$-string theory. 

It is known that the 
${1\over 2}K_3$ can be obtained by blowing up 9 points of ${\bf CP}^2$ and
is also called as an almost del Pezzo surface ${\cal B}_9$. 
Elements in the second homology group of ${1\over 2}K_3$
span a unimodular lattice $\Gamma^{9,1}$ with signature $(1,9)$ which contains the
lattice $\Gamma^8$ of $E_8$ , $\Gamma^{9,1}=\Gamma^{8}\oplus \Gamma^{1,1}$. 
The existence of the lattice $\Gamma^8$ 
is the origin of the $E_8$ symmetry in $E$-string theory.

It is known that the del Pezzo surface ${\cal B}_9$ possesses an elliptic fibration 
over a ${\bf P}^1$ with 12 singular fibers.
The fibration
is described by a family of elliptic curves written in a Weierstrass form
\be
{\cal C}: \hskip2mm y^2=4x^3-f(u,\tau;m_i)x-g(u,\tau;m_i),\hskip20mm i=1,2,\ldots ,8
\label{swcurve}\ee 
where the coefficient functions $f,g$ are degree 4 and 6 polynomials in $u$, 
respectively
\ba
&&f(u,\tau;m_i)=a_0(\tau)u^4+a_1(\tau;m_i)u^3+a_2(\tau;m_i)u^2+a_3(\tau;m_i)u+
a_4(\tau;m_i),
\label{ai}\\
&&g(u,\tau;m_i)=
b_0(\tau)u^6+b_1(\tau;m_i)u^5+b_2(\tau;m_i)u^4+b_3(\tau;m_i)u^3+b_4(\tau;m_i)u^2 
\no \\
&&
\hskip50mm +b_5(\tau;m_i)u+b_6(\tau;m_i).\label{bi}
\ea
$u$ parametrizes the base ${\bf P}^1$ and $m_i\hskip1mm i=1,2,\ldots ,8$
represent the position of the blow-up points (position of the 9th blow-up
point is fixed by the other 8 points).
$\tau$ gives the modulus of the elliptic fiber at $u=\infty$ and is also the 
modulus of
the torus $T^2$ upon which $E$-string theory is compactified down to 4-dimensions. 
$\{m_i\}$ correspond physically to the masses of $8$ hypermultiplets
in ${\cal N}=2$ $SU(2)$ gauge theory on ${\bf R}^4\times T^2$.

With $\tau$ being the modulus at $u=\infty$, $a_0(\tau),b_0(\tau)$ are fixed to
\be
a_0(\tau)={1\over 12} E_4(\tau),\hskip4mm b_0(\tau)={1\over 216}E_6(\tau),
\ee
where $E_n$ is the Eisenstein series with weight $n$,
\be
E_{2n}(\tau)=1+{(2\pi i)^{2n}\over (2n-1)!\,\zeta(2n)}
\sum_{m=1}\sigma_{2n-1}(m)q^m, \hskip3mm \sigma_k(m)=\sum_{d|m}d^k,
\hskip3mm q=e^{2\pi i\tau}. 
\ee
By shifting the variable $u$ we can eliminate either one of the functions
$a_1$ or $b_1$. In the following we choose a ``gauge'' where we put
\be
a_1(\tau;m_i)=0.
\ee  

We call eq.(\ref{swcurve}) as the six-dimensional Seiberg--Witten (SW) curve:
the six-dimensional curve determines the prepotential of $E$-string theory and 
gives a 
complete description of its low-energy dynamics.
Previously there were attempts at deriving the six-dimensional curve 
\cite{Ga,GMS,MNWa,Moh},
however, only partial results with a few non-vanishing mass parameters $\{m_i\}$ 
were
obtained. In this paper we will determine all the coefficient functions 
$a_j,b_j$ and the curve (\ref{swcurve})
for arbitrary values of $\{m_i\}$. It turns out that the functions $a_j,b_j$
are expressed in terms of characters of affine $E_8$ at level 
$j$.

Six-dimensional curves with lower-rank symmetry groups 
can be easily obtained from (\ref{swcurve}) by taking some of the mass 
parameters $\{m_i\}$ to $\infty$ or adjusting them to special values.
We shall show that the known 
five-dimensional SW curves \cite{MNWa} are reproduced from 
the six-dimensional curve
by taking the limit $\mbox{Im}\tau\rightarrow +\infty$. 
Four-dimensional curves \cite{SW,MNa,MNb} are also obtained by further 
taking the remaining period of $T^2$ to 0. Thus our six-dimensional curve 
serves as some kind of master theory encompassing all possible SW theories.

Let us now recall the relation between BPS states of the $E$-string and  
holomorphic curves in the Calabi--Yau 3-fold \cite{KMV} 
and also the partition function of the ${\cal N}=4$ $U(n)$ 
Yang--Mills theory on ${1\over 2}K_3$ \cite{MNVW}.
Consider an F-theory compactification down to 6-dimensions 
on a Calabi--Yau 3-fold $K$ which is elliptically fibered
over a base $B$. We choose a curve $C$ inside the base $B$ so that the elliptic 
fibration 
restricted to $C$ gives the ${1\over 2}K_3$. Consider a D3-brane wrapped around
$C$, which gives rise to a string in 6-dimensions. 
Low energy dynamics of such a string can be studied by looking at its BPS spectrum. 
We compactify the 5th dimension on a circle $R$ and count the BPS states of the 
string 
with winding number $n$ and momentum $k$. We denote the number of these states as 
$Z_{n,k}$.

Next by using a duality between F- and M-theory we go to a M-theory description:
a compactification of F-theory on $K\times S^1$, where the $S^1$ has radius $R$, 
is equivalent to
a compactification of M-theory on $K$ with the K\"ahler parameter of the elliptic 
fiber being 
equal to $1/R$. In M-theory 5-dimensional BPS states are obtained by wrapping a 
membrane
around holomorphic curves. Then the number of BPS states $Z_{n,k}$ is
given by the number
of curves in the class $n[C]+k[E]$ where $[C]$ ($[E]$) denotes the class of the 
base 
(elliptic fiber) of ${1\over 2}K_3$.
Thus the counting of BPS states of 6-dimensional $E$-string wrapped around a 
circle  
is related to the counting of holomorphic curves in $K$ which can be analyzed by 
using the technique of mirror symmetry.

If we introduce M5-branes, the number of BPS states becomes further related to the 
partition 
function of $U(n)$ 
Yang--Mills theory on ${1\over 2}K_3$ \cite{MNVW}. Consider a M5-brane wrapped 
around 
${1\over 2}K_3$. One then obtains a string in 5-dimensions. In order to study its 
spectrum we
may wrap the string around a circle and compute its toroidal partition 
function. 
Then 
the M5-brane
becomes effectively wrapped around ${1\over 2}K_3\times T^2$. If we consider the 
string
$n$-times wound around a circle, M5-brane wraps around $T^2$ $n$-times and one obtains 
${\cal N}=4$ $U(n)$  
Yang--Mills theory on ${1\over 2}K_3$.
The gauge coupling constant is given by the modulus $\tau$ of the torus and 
the momentum $k$ is mapped to the instanton number.
Thus $Z_{n,k}$ is also interpreted as the $k$-instanton contribution to the 
partition 
function
of $U(n)$ gauge theory on ${1\over 2}K_3$.

We want to determine the prepotential of the theory
\ba
{\cal F}(\phi,\tau;m_i)={\cal F}_{classical}-{i\over 8\pi^3}
\sum_{n= 1}\,q^{n/2}Z_n(\tau;m_i)\,e^{2\pi i\,n\phi}, \hskip2mm q^{n/2}
Z_n(\tau;m_i)=\sum_{k=0}Z_{n,k}(m_i)q^k,\no\\
 \hskip2mm q=e^{2\pi i\tau}\hspace{5mm}
\ea
where $\phi$ is the K\"ahler parameter which is dual (mirror) to the variable $u$.
We have inserted a factor $q^{n/2}$ for the sake of a good
modular property of $Z_n$:
$Z_n$ becomes a (quasi-)Jacobi form of weight $-2$.

Actually the functions $Z_n$ have already been computed recursively up to
$Z_4$ by making use of the holomorphic anomaly and the gap condition
($Z_{n,k}=0$ for $0<k<n$) in \cite{MNVW}. One may continue this computation and
obtain more data on $Z_n$.
On the other hand, from the SW
curve (\ref{swcurve}) one can compute the prepotential by making use of the 
mirror-type transformation
and express $Z_n$ in terms of the 
coefficient functions
$a_j,b_j$. It turns out that $Z_n$ is written as a polynomial 
in $a_j,b_j,\,0\le j\le n$ and in particular linear in $a_n,b_n$. Hence given the data 
$\{Z_n\}$ it is easy to determine $\{a_j,b_j\}$. In our actual computation we have
used the data up to $Z_{8}$ to determine the Seiberg--Witten curve.

\subsection{Instanton Expansion}

 Let us next describe the standard mirror-type transformation
in Seiberg--Witten theory adopted to the 
present situation \cite{MNWb}.
The coupling constant $\tilde{\tau}$ of ${\cal N}=2$ gauge theory is given by the 
modulus
of the elliptic curve
\be
j(\tilde{\tau})={1728f(u,\tau;m_i)^3\over f(u,\tau;m_i)^3-27g(u,\tau;m_i)^2}.
\label{couplinga}\ee
Here $j$ is the elliptic j-function. We may expand the right-hand side of
the above equation in $1/u$ and obtain
\ba
j(\tilde{\tau})&=&{1728E_4(\tau)^3\over E_4(\tau)^3-E_6(\tau)^2}+{1\over 4}
{E_4(\tau)^3E_6(\tau)b_1(\tau;m_i)\over \Delta(\tau)^{2}}{1\over u}+\cdots
\no\\
&=&
j(\tau)+{1\over 4}{E_4(\tau)^3E_6(\tau)b_1(\tau;m_i)\over \Delta(\tau)^{2}}
{1\over u}+\cdots,
\label{couplingb}\ea
where $\Delta=({E_4}^3-{E_6}^2)/1728=\eta^{24}$.
Thus $\tilde{\tau}=\tau$ at $u=\infty$. By inverting the j-function 
in (\ref{couplingb})
we can generate 
a Taylor series expansion of ${\tilde{\tau}}$ around $u=\infty$,
\be
\tilde{\tau}=\tau+{i\over 8\pi}{E_4b_1\over \Delta}{1\over u}
+{i \over 384\pi}{-4E_6\Delta a_2+48E_4\Delta b_2+5E_4E_6{b_1}^2+E_2{E_4}^2
{b_1}^2\over 
\Delta^2}{1\over u^2}+\cdots.
\ee
Periods of the torus $\omega_1,\omega_2$ are given by
\be
\omega_1={i\over 2\pi}\left({E_4(\tilde{\tau})\over 12f(u,\tau;m_i)}
\right)^{1/4},\hskip3mm \omega_2=\tilde{\tau}\omega_1.
\ee 
Both of these $\omega_i$ are also expanded in Taylor series in $1/u$. 
We can then easily integrate them
and obtain functions $\phi,\phi_D$
\be
\phi=\int du \,\omega_1(u,\tau;m_i),\hskip3mm 
\phi_D=\int du\,\omega_2(u,\tau;m_i)=\int du \,\tilde{\tau}(u,\tau;m_i)
\omega_1(u,\tau;m_i).
\ee
Since
$\partial \phi/\partial u=\omega_1, \,
\partial \phi_D/ \partial u=\omega_2$,
$\phi,\,\phi_D$ correspond to the variables of the Coulomb branch
$a,\, a_D$ in the 4-dimensional Seiberg--Witten
theory. Prepotential is defined by
\be
{\partial {\cal F}\over \partial \phi}=\phi_D
\ee
as usual. 

Lower order computation goes as follows: $\omega_1$ has an expansion
\ba
&&\omega_1={i \over 2\pi}{1\over u}-{i \over 96\pi}{(E_2E_4-E_6)b_1\over \Delta}
{1\over u^2}+\cdots.
\ea
$\phi$ is then given by
\be
\phi=\phi_0+{i\over 2\pi}\ln u+{i\over 96\pi}{(E_2E_4-E_6)b_1\over \Delta}{1\over u}
+\cdots
\ee
where $\phi_0$ is an integration constant.
Thus 
\be
{1\over u}=e^{2\pi i(\phi-\phi_0)}+{1\over 48}
{(E_2E_4-E_6)b_1\over \Delta}e^{4\pi i(\phi-\phi_0)}+\cdots
\ee
and therefore
\ba
{\partial^2 {\cal F}\over \partial\phi^2}=
\tilde{\tau}=\tau\Eqn{+}{i\over 8\pi}{E_4b_1\over \Delta}e^{2\pi i(\phi-\phi_0)}
\no\\
\Eqn{+}{i \over 192\pi}
{-2E_6\Delta a_2+24E_4\Delta b_2+2E_4E_6{b_1}^2+E_2{E_4}^2{b_1}^2\over 
\Delta^2}e^{4\pi i (\phi-\phi_0)}+\cdots.\hspace{10mm}
\ea
By integrating twice in $\phi$ we obtain the prepotential. 
In the following we choose $\phi_0=-(\ln \eta^{12})/2\pi i-1/2$.

We have performed the above transformation up to higher orders
and obtained the partition functions 
expressed in terms of the coefficient functions of the SW curve. 
We present lower order terms
\ba
Z_1\Eqn{=}-\frac{1}{4}\,
\frac{E_4}{\eta^{12}}\,{b_1}
\ ,\label{z1a}\\
Z_2\Eqn{=}{1\over 8}
\left(-\frac{1}{24}\,{E_6}\,{a_2}
+\frac{1}{2} \,{E_4}\,{b_2}
+\frac{1}{24}\,\frac{{E_4}\,{E_6}}{\eta^{24}}\,{b_1}^{2}
+\frac{1}{48}\,\frac{{E_2}\,{E_4}^{2}}{\eta^{24}}\,{b_1}^{2}\right)
\ ,\label{z2a}\\
Z_3\Eqn{=}{1\over 27}
\left( \frac{1}{16} \,\eta^{12}\,{E_6}\,{a_3}
-\frac{3}{4}  \,\eta^{12}\,{E_4}\,{b_3}
-\frac{9}{2}  \,\eta^{12}\,{a_2}\,{b_1}
+\frac{5}{512}\,\frac{{E_4}^{3}}{\eta^{12}}\,{a_2}\,{b_1}\right.
\no\\
&&\left.
+\frac{3}{512} \,\frac{{E_2}\,{E_4}\,{E_6}}{\eta^{12}}\,{a_2}\,{b_1}
-\frac{15}{128}\,\frac{{E_4}\,{E_6}}{\eta^{12}}\,{b_2}\,{b_1}
-\frac{9}{128} \,\frac{{E_2}\,{E_4}^{2}}{\eta^{12}}\,{b_2}\,{b_1}\right.
\no\\
&&\left.
+\frac{591}{64} \,\frac{{E_4}}{\eta^{12}}\,{b_1}^{3}
-\frac{17}{2048}\,\frac{{E_4}^{4}}{\eta^{36}}\,{b_1}^{3}
-\frac{3}{512}  \,\frac{{E_2}\,{E_4}^{2}\,{E_6}}{\eta^{36}}\,{b_1}^{3}
-\frac{3}{2048} \,\frac{{E_2}^{2}\,{E_4}^{3}}{\eta^{36}}\,{b_1}^{3}\right)
\ .\label{z3a}
\ea
We have obtained similar formulas up to $Z_8$.

\subsection{Holomorphic Anomaly and Gap Condition}

Partition functions $Z_n$ of $U(n)$ gauge theory on ${1\over 2}K_3$
have been computed in \cite{MNVW}
by imposing the relation of holomorphic anomaly and gap condition (see 
\cite{Yo} for a different analysis).
Holomorphic anomaly occurs in the gauge theory 
due to the appearance of reducible connections and corresponds to the
existence of threshold bound states in the $E$-string spectrum.
Holomorphic anomaly enters via the dependence on $E_2$ 
of $Z_n$ and governed by the relation \cite{MNWb}
\be
{\partial Z_n\over \partial E_2}
={1 \over 24}\sum_{m=1}^{n-1}m(n-m)Z_mZ_{n-m}\ .
\label{anomaly}\ee
(\ref{anomaly}) corresponds to the reduction of the $U(n)$ connection down 
to that of $U(m)\times U(n-m)$. 
Holomorphic anomaly is tied with the fact that ${1\over 2}K_3$ does not possess
a holomorphic 2-form and 
there exists no mass perturbation of ${\cal N}=4$ theory
which splits the location of $n$ M5-branes. Reduction of supersymmetry
leads to a binding force among the branes.

Gap condition on the other hand 
requires that instantons with degree $k<n$ do not exist in
$U(n)$ gauge theory. This comes from the positivity of intersection
numbers among holomorphic curves in ${1\over 2}K_3$.

These two conditions and consideration of modular invariance
uniquely determine the amplitudes.
We list the first 3 instanton amplitudes rewritten slightly from \cite{MNVW}
\ba
Z_1\Eqn{=}\frac{P(m_i;\tau)}{\eta^{12}}\label{z1b}\\
\mbox{where}
&&P(m_i;\tau)={1\over
2}\left[\sum_{\ell=1}^4\prod_{j=1}^8\theta_{\ell}(m_j|\tau)\right],\\
\no\\
Z_2\Eqn{=}\frac{1}{\eta^{24}}\biggl[
f_{20}(\tau) P(2 m_i;2\tau)
+f_{21}(\tau) P\Bigl(m_i;\frac{\tau}{2}\Bigr)
+f_{21}(\tau+1) P\Bigl(m_i;\frac{\tau+1}{2}\Bigr)\biggr]
+\frac{1}{24}E_2{Z_1}^2\no\\
\label{z2b}\\
\mbox{where}
&&f_{20}(\tau)=\frac{1}{24}\theta_3(\tau)^4\theta_4(\tau)^4
  \left(\theta_3(\tau)^4+\theta_4(\tau)^4\right),\\
&&f_{21}(\tau)=-\frac{1}{384}\theta_3(\tau)^4\theta_2(\tau)^4
  \left(\theta_3(\tau)^4+\theta_2(\tau)^4\right),\\
\no\\
Z_3\Eqn{=}\frac{1}{\eta^{36}}\biggl[
f_{30}(\tau) P(3 m_i;3\tau)
+f_{31}(\tau) P\Bigl(m_i;\frac{\tau}{3}\Bigr)\no\\
&&\hspace{10mm}
+f_{31}(\tau+1) P\Bigl(m_i;\frac{\tau+1}{3}\Bigr)
+f_{31}(\tau+2) P\Bigl(m_i;\frac{\tau+2}{3}\Bigr)
-\frac{1}{288}E_4(\tau) P(m_i;\tau)^3\biggr]\no\\
&&+\frac{1}{6}E_2Z_2Z_1-\frac{1}{288} {E_2}^2{Z_1}^3\label{z3b}\\
\no\\
\mbox{where}
&&f_{30}(\tau)=
  \frac{5}{216}\frac{\eta(\tau)^{36}}{\eta(3\tau)^{12}}
  +\frac{9}{8}\eta(\tau)^{24},\\
&&f_{31}(\tau)=
  \frac{5}{24}\frac{\eta(\tau)^{36}}{\eta(\tau/3)^{12}}
  +\frac{1}{72}\eta(\tau)^{24}.
\ea

Note that the function $P(m_i;\tau)$ is (up to $\eta^8$) the character of the level-one
representation of $\widehat{E}_8$ and $P(2m_i;2\tau),P(m_i;{\tau\over 2}),
P(m_i;{\tau+1\over 2})$ are linear combinations of characters of
3 level-two representations of $\widehat{E}_8$. In general the amplitude
$Z_n$ is expressed
in terms of level-$n$ representations of $\widehat{E}_8$. There exist 1,3,5,10,15,27 
distinct representations of $\widehat{E}_8$ at the levels 1,2,3,4,5,6, respectively.
Coefficient functions $f_{n0}$ are modular forms of $\Gamma_1(n)$ which consists
of matrices of the form $\Biggl(\begin{array}{cc}a & b\\ c& d\end{array}\Biggr)$
with $ad-bc=1$ and $a,d=1,c=0 \hskip1mm \mbox{mod}\,n$. Functions
$f_{n1}$ are the S-transform of $f_{n0}$.

Using holomorphic anomaly and the gap condition we have obtained the
partition functions up to $Z_8$.

\subsection{Six-dimensional Curve}

Now by using these data  
we can determine the coefficient functions of SW curve.
We present the first few terms and relegate the rest to
\ref{app6dim}
\ba
b_1\Eqn{=}-4{P(m_i;\tau)\over E_4},\\
\no\\
a_2\Eqn{=}\frac{1}{E_4\Delta}\biggl[
f_{a2,0}(\tau) P(2 m_i;2\tau)
+f_{a2,1}(\tau) P\Bigl(m_i;\frac{\tau}{2}\Bigr)
+f_{a2,1}(\tau+1) P\Bigl(m_i;\frac{\tau+1}{2}\Bigr)\biggr]\\
\mbox{where}
&&f_{a2,0}(\tau)=\frac{2}{3}\left(E_4(\tau)-9\theta_2(\tau)^8\right),\\
&&f_{a2,1}(\tau)=\frac{1}{24}\left(E_4(\tau)-9\theta_4(\tau)^8\right),\\
\no\\
b_2\Eqn{=}\frac{1}{{E_4}^2\Delta}\biggl[
f_{b2,0}(\tau) P(2 m_i;2\tau)
+f_{b2,1}(\tau) P\Bigl(m_i;\frac{\tau}{2}\Bigr)
+f_{b2,1}(\tau+1) P\Bigl(m_i;\frac{\tau+1}{2}\Bigr)\biggr]\\
\mbox{where}
&& f_{b2,0}(\tau)=
  \frac{1}{36}\left(\theta_3(\tau)^4+\theta_4(\tau)^4\right)
  \left(E_4(\tau)^2+60 E_4(\tau)\theta_2(\tau)^8
  -45\theta_2(\tau)^{16}\right),\\
&& f_{b2,1}(\tau)=
  -\frac{1}{576}\left(\theta_3(\tau)^4+\theta_2(\tau)^4\right)
  \left(E_4(\tau)^2+60 E_4(\tau)\theta_4(\tau)^8
  -45\theta_4(\tau)^{16}\right),\\
\no\\
a_3\Eqn{=}\frac{1}{E_4^2\Delta^2}\biggl[
f_{a3,0}(\tau) P(3 m_i;3\tau)
+f_{a3,1}(\tau) P\Bigl(m_i;\frac{\tau}{3}\Bigr)
+f_{a3,1}(\tau+1) P\Bigl(m_i;\frac{\tau+1}{3}\Bigr)\no\\
&&\hspace{15mm}
+f_{a3,1}(\tau+2) P\Bigl(m_i;\frac{\tau+2}{3}\Bigr)
+\frac{2}{3}E_6(\tau) P(m_i;\tau)^3\biggr],\\
\mbox{where}
&& f_{a3,0}(\tau)=\frac{1}{3}E_4(\tau) h_2(\tau)^2
  \left(7 E_4(\tau)-9 h_0(\tau)^4\right),\\
&& f_{a3,1}(\tau)=-\frac{1}{3^8}E_4(\tau) h_3(\tau/3)^2
  \left(7 E_4(\tau)-h_0(\tau/3)^4\right),\\
\no\\
b_3\Eqn{=}\frac{1}{{E_4}^3\Delta^2}\biggl[
f_{b3,0}(\tau) P(3 m_i;3\tau)
+f_{b3,1}(\tau) P\Bigl(m_i;\frac{\tau}{3}\Bigr)
+f_{b3,1}(\tau+1) P\Bigl(m_i;\frac{\tau+1}{3}\Bigr)\no\\
&&\hspace{15mm}
+f_{b3,1}(\tau+2) P\Bigl(m_i;\frac{\tau+2}{3}\Bigr)
+\frac{1}{54}\left(8{E_4}^3-5{E_6}^2\right) P(m_i;\tau)^3\biggr],\\
\mbox{where}
&&f_{b3,0}(\tau)=\frac{1}{18}E_4(\tau)^2 h_2(\tau)^2
  \left(32 h_2(\tau)^2+48h_2(\tau) h_0(\tau)^3-81 h_0(\tau)^6\right),\\
&&f_{b3,1}(\tau)=\frac{1}{2\cdot 3^{12}} E_4(\tau)^2 h_3(\tau/3)^2
  \left(32 h_3(\tau/3)^2+48 h_3(\tau/3)h_0(\tau/3)^3
  -81h_0(\tau/3)^6\right).\no\\
\ea
Here functions $h_i$ are defined by
\ba
h_0(\tau)\Eqn{=}\sum_{n_1,n_2=-\infty}^{\infty}q^{n_1^2+n_2^2-n_1n_2}
=\theta_3(2\tau)\theta_3(6\tau)+\theta_2(2\tau)\theta_2(6\tau),\\
h_2(\tau)\Eqn{=}\frac{\eta(\tau)^9}{\eta(3\tau)^3},\hskip3mm 
h_3(\tau)=27\frac{\eta(3\tau)^9}{\eta(\tau)^3}.
\ea

It is fortunate that the six-dimensional curve can be written in a relatively 
compact expression. As we have mentioned, coefficient functions $a_j,b_j$
are expressed in terms of affine $E_8$ characters at level-$j$.
Functions $f_{*n,0}$ are modular forms of $\Gamma_1(n)$ and
$f_{*n,1}$ are their S-transform.
We have checked that our curve reduces to the results
of \cite{MNWa} when six of the mass parameters are set equal to zero.

\subsection{Five-dimensional Curve}

\begin{figure}[t]
\begin{center}
\unitlength=2.4pt
\begin{picture}(70,30)
\put( 0,10){\circle{2}}
\put( 0,10){${\!}^2$}
\put( 0, 6){${\!\!\!\!\>}_{(1)}$}
\put( 1,10){\line(1,0){8}}
\put(10,10){\circle{2}}
\put(10,10){${\!}^4$}
\put(10, 6){${\!\!\!\!\>}_{(3)}$}
\put(11,10){\line(1,0){8}}
\put(20,10){\circle{2}}
\put(20,10){${\,}^6$}
\put(20, 6){${\!\!\!\!\>}_{(4)}$}
\put(21,10){\line(1,0){8}}
\put(30,10){\circle{2}}
\put(30,10){${\!}^5$}
\put(30, 6){${\!\!\!\!\>}_{(5)}$}
\put(31,10){\line(1,0){8}}
\put(40,10){\circle{2}}
\put(40,10){${\!}^4$}
\put(40, 6){${\!\!\!\!\>}_{(6)}$}
\put(41,10){\line(1,0){8}}
\put(50,10){\circle{2}}
\put(50,10){${\!}^3$}
\put(50, 6){${\!\!\!\!\>}_{(7)}$}
\put(51,10){\line(1,0){8}}
\put(60,10){\circle{2}}
\put(60,10){${\!}^2$}
\put(60, 6){${\!\!\!\!\>}_{(8)}$}
%\put(61,10){\line(1,0){8}}
%\put(70,10){\circle{2}}
%\put(70,10){${\!}^1$}
%\put(70, 6){${\!\!\!\!\>}_{(0)}$}
\put(20,11){\line(0,1){8}}
\put(20,20){\circle{2}}
\put(20,20){${\!}^3$}
\put(20,20){${\hspace{0.4em}}_{(2)}$}
\end{picture}
\caption{Dynkin diagram for $E_8$:
numbers attached to nodes denote the levels of representations
and the numbers in parentheses show their labels.
\label{aE_8diagram}}
\end{center}
\end{figure}
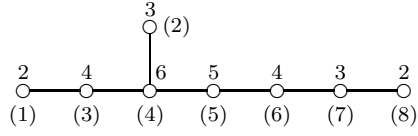

Let us next see how our result reproduces the five-dimensional curve 
in the limit $q\rightarrow 0$. Affine $E_8$ characters are reduced to
those of the representations of finite-dimensional $E_8$ algebra in this limit.
Let us first introduce a labeling of the fundamental representations
of $E_8$ 
as in Fig. \ref{aE_8diagram} where we also show the level of each 
representation. 
We then introduce a notation 
\be
\rho=\Biggl[\begin{array}{ccccccc} & & n_2 & & & & \\
n_1 &n_3 &n_4 &n_5 &n_6&n_7&n_8\end{array}\Biggr]=\sum_{i=1}^8 n_i\rho_i,
\hskip4mm n_i=0,1,2,\ldots
\ee
for a representation $\rho$ of $E_8$ whose highest weight is given by the sum of
highest weights of the fundamental representation $\rho_i$ with multiplicity 
$n_i$. The level of the representation $\rho$ is given by $\sum_i \ell_in_i$ 
where $\ell_i$ is the level of $\rho_i$. Weyl-orbit character 
for the representation $\rho$ is given by
\be
w_{\rho}(m_i)=\sum_{\vec{\alpha}\in \Lambda(\rho)}e^{i\vec{m}\cdot\vec{\alpha}}
\ee
where $\vec{\alpha}$ runs over the weights $\Lambda(\rho)$ of the $E_8$ 
Weyl orbit containing the highest-weight state of the representation $\rho$.

With this preparation we can present the five-dimensional $E_8$ curve 
\ba
&&y^2=4x^3-F(u;m_i)x-G(u;m_i),\\
&&F(u;m_i)=A_0u^4+A_2(m_i)u^2+A_3(m_i)u+A_4(m_i),\no\\
&&G(u;m_i)=B_0u^6+B_1u^5+B_2(m_i)u^4+B_3(m_i)u^3+B_4(m_i)u^2
+B_5(m_i)u+B_6(m_i)\no
\ea
with
\ba
A_0\Eqn{=}\tfrac{1}{12}, \hskip3mm
A_2=
- \tfrac{2}{3}         \WOEei{1}{0}{0}{0}{0}{0}{0}{0}%2 - 4
+ \tchara{12}          \WOEei{0}{0}{0}{0}{0}{0}{0}{1}%2 - 2
- \tchara{1440},\\
&&\no\\
A_3\Eqn{=}
- \tchara{2}           \WOEei{0}{1}{0}{0}{0}{0}{0}{0}%3 - 8
%+ 0                   \WOEei{0}{0}{0}{0}{0}{0}{1}{0}%3 - 6
+ \tchara{96}          \WOEei{1}{0}{0}{0}{0}{0}{0}{0}%2 - 4
- \tchara{1152}        \WOEei{0}{0}{0}{0}{0}{0}{0}{1}%2 - 2
+ \tchara{103680},\\
&&\no \\
A_4\Eqn{=}
  \tfrac{4}{3}         \WOEei{2}{0}{0}{0}{0}{0}{0}{0}%4 - 16
- \tfrac{4}{3}         \WOEei{0}{0}{1}{0}{0}{0}{0}{0}%4 - 14
- \tfrac{8}{3}         \WOEei{0}{0}{0}{0}{0}{1}{0}{0}%4 - 12
- \tfrac{16}{3}        \WOEei{1}{0}{0}{0}{0}{0}{0}{1}%4 - 10
+ \tchara{24}          \WOEei{0}{0}{0}{0}{0}{0}{0}{2}%4 - 8
\no \\
&&
+ \tfrac{328}{3}       \WOEei{0}{1}{0}{0}{0}{0}{0}{0}%3 - 8
+ \tchara{48}          \WOEei{0}{0}{0}{0}{0}{0}{1}{0}%3 - 6
- \tfrac{8960}{3}      \WOEei{1}{0}{0}{0}{0}{0}{0}{0}%3 - 4
+ \tchara{28128}       \WOEei{0}{0}{0}{0}{0}{0}{0}{1}%2 - 2
- \tchara{2105280},
\\
&&\no\\
B_0\Eqn{=}\tfrac{1}{216},\hskip3mm B_1=-\tchara{4},
\hskip3mm B_2=
- \tfrac{1}{18}        \WOEei{1}{0}{0}{0}{0}{0}{0}{0}%2 - 4
- \tchara{3}           \WOEei{0}{0}{0}{0}{0}{0}{0}{1}%2 - 2
+ \tchara{840},\\
&&\no\\
B_3\Eqn{=}
- \tfrac{1}{6}         \WOEei{0}{1}{0}{0}{0}{0}{0}{0}%3 - 8
- \tchara{4}           \WOEei{0}{0}{0}{0}{0}{0}{1}{0}%3 - 6
- \tchara{8}           \WOEei{1}{0}{0}{0}{0}{0}{0}{0}%2 - 4
+ \tchara{528}         \WOEei{0}{0}{0}{0}{0}{0}{0}{1}%2 - 2
- \tchara{79680},\\
&&\no\\
B_4\Eqn{=}
  \tfrac{2}{9}         \WOEei{2}{0}{0}{0}{0}{0}{0}{0}%4 - 16
+ \tfrac{1}{9}         \WOEei{0}{0}{1}{0}{0}{0}{0}{0}%4 - 14
- \tfrac{28}{9}        \WOEei{0}{0}{0}{0}{0}{1}{0}{0}%4 - 12
- \tfrac{152}{9}       \WOEei{1}{0}{0}{0}{0}{0}{0}{1}%4 - 10
- \tchara{92}          \WOEei{0}{0}{0}{0}{0}{0}{0}{2}%4 - 8
\no\\
&&
- \tfrac{316}{9}       \WOEei{0}{1}{0}{0}{0}{0}{0}{0}%3 - 8
+ \tchara{416}         \WOEei{0}{0}{0}{0}{0}{0}{1}{0}%3 - 6
+ \tfrac{13088}{9}     \WOEei{1}{0}{0}{0}{0}{0}{0}{0}%2 - 4
- \tchara{35536}       \WOEei{0}{0}{0}{0}{0}{0}{0}{1}%2 - 2
+ \tchara{3911520},\\
&&\no\\
B_5\Eqn{=}
  \tfrac{2}{3}         \WOEei{1}{1}{0}{0}{0}{0}{0}{0}%5 - 22
- \tfrac{2}{3}         \WOEei{0}{0}{0}{0}{1}{0}{0}{0}%5 - 20
- \tfrac{16}{3}        \WOEei{1}{0}{0}{0}{0}{0}{1}{0}%5 - 18
- \tfrac{74}{3}        \WOEei{0}{1}{0}{0}{0}{0}{0}{1}%5 - 16
- \tchara{96}          \WOEei{0}{0}{0}{0}{0}{0}{1}{1}%5 - 14
- \tfrac{160}{3}       \WOEei{2}{0}{0}{0}{0}{0}{0}{0}%4 - 16
\no\\
&&
- \tfrac{280}{3}       \WOEei{0}{0}{1}{0}{0}{0}{0}{0}%4 - 14
+ \tchara{80}          \WOEei{0}{0}{0}{0}{0}{1}{0}{0}%4 - 12
+ \tchara{1088}        \WOEei{1}{0}{0}{0}{0}{0}{0}{1}%4 - 10
+ \tchara{7872}        \WOEei{0}{0}{0}{0}{0}{0}{0}{2}%4 - 8
+ \tfrac{9680}{3}      \WOEei{0}{1}{0}{0}{0}{0}{0}{0}%3 - 8
\no \\
&&
- \tchara{15264}       \WOEei{0}{0}{0}{0}{0}{0}{1}{0}%3 - 6
- \tfrac{196448}{3}    \WOEei{1}{0}{0}{0}{0}{0}{0}{0}%2 - 4
+ \tchara{1075776}     \WOEei{0}{0}{0}{0}{0}{0}{0}{1}%2 - 2
- \tchara{97251840},\\
&&\no\\
B_6\Eqn{=}
- \tfrac{8}{27}        \WOEei{3}{0}{0}{0}{0}{0}{0}{0}%6 - 36
+                      \WOEei{0}{2}{0}{0}{0}{0}{0}{0}%6 - 32
+ \tfrac{4}{9}         \WOEei{1}{0}{1}{0}{0}{0}{0}{0}%6 - 32
+ \tfrac{2}{9}         \WOEei{0}{0}{0}{1}{0}{0}{0}{0}%6 - 30
%+ 0                   \WOEei{1}{0}{0}{0}{0}{1}{0}{0}%6 - 28
- \tfrac{16}{9}        \WOEei{2}{0}{0}{0}{0}{0}{0}{1}%6 - 26
- \tfrac{8}{3}         \WOEei{0}{1}{0}{0}{0}{0}{1}{0}%6 - 26
\no\\
&&
- \tchara{8}           \WOEei{0}{0}{0}{0}{0}{0}{2}{0}%6 - 24
- \tfrac{88}{9}        \WOEei{0}{0}{1}{0}{0}{0}{0}{1}%6 - 24
- \tfrac{112}{3}       \WOEei{0}{0}{0}{0}{0}{1}{0}{1}%6 - 22
- \tfrac{1000}{9}      \WOEei{1}{0}{0}{0}{0}{0}{0}{2}%6 - 20
- \tfrac{992}{3}       \WOEei{0}{0}{0}{0}{0}{0}{0}{3}%6 - 18
\no\\
&&
- \tchara{70}          \WOEei{1}{1}{0}{0}{0}{0}{0}{0}%5 - 22
- \tchara{64}          \WOEei{0}{0}{0}{0}{1}{0}{0}{0}%5 - 20
+ \tfrac{112}{9}       \WOEei{1}{0}{0}{0}{0}{0}{1}{0}%5 - 18
+ \tchara{636}         \WOEei{0}{1}{0}{0}{0}{0}{0}{1}%5 - 16
+ \tchara{3472}        \WOEei{0}{0}{0}{0}{0}{0}{1}{1}%5 - 14
\no\\
&&
+ \tchara{1584}        \WOEei{2}{0}{0}{0}{0}{0}{0}{0}%4 - 16
+ \tchara{3038}        \WOEei{0}{0}{1}{0}{0}{0}{0}{0}%4 - 14
+ \tfrac{7376}{9}      \WOEei{0}{0}{0}{0}{0}{1}{0}{0}%4 - 12
- \tchara{18480}       \WOEei{1}{0}{0}{0}{0}{0}{0}{1}%4 - 10
- \tchara{176608}      \WOEei{0}{0}{0}{0}{0}{0}{0}{2}%4 - 8
\no\\
&&
- \tfrac{196384}{3}    \WOEei{0}{1}{0}{0}{0}{0}{0}{0}%3 - 8
+ \tchara{197968}      \WOEei{0}{0}{0}{0}{0}{0}{1}{0}%3 - 6
+ \tchara{936200}      \WOEei{1}{0}{0}{0}{0}{0}{0}{0}%2 - 4
- \tchara{12291232}    \WOEei{0}{0}{0}{0}{0}{0}{0}{1}%2 - 2
+ \tchara{971250560}.
\no\\
&&
\ea
(It is also possible to represent this curve using Weyl-orbit characters only
for fundamental representations of $E_8$.
See \ref{app5dim}
In this case products of characters appear
in the formula.)

This curve has a manifest $E_8$ symmetry and can be derived from the 
expression 
of \cite{MNWa} with $SO(16)$ symmetry by a 
suitable shifting of variables and 
rearrangement of terms.
Note that $A_j,B_j$ contain representations with levels $1\le l \le j$
(constant term is interpreted as level-one representation).
This is consistent with the fact that an affine representation of 
level $j$ is reduced to finite-dimensional representations with levels
$l\le j$ in the limit $q\to 0$.

In fact we can explicitly show that
\ba
\lim_{q\to 0}a_j(\tau;m_i)\Eqn{=}A_j(m_i),\quad j=0,1,2,\ldots,4,\\
\label{blimit}
\lim_{q\to 0}b_j(\tau;m_i)\Eqn{=}B_j(m_i),\quad j=0,1,2,\ldots,6
\ea
hold. Since
\be
P(m_i;\tau),\ E_4(\tau)\to 1\quad \mbox{as}\quad q\to 0
\ee
we can easily check the relation (\ref{blimit}) for $b_1$.
In general it is convenient to use the basis of affine Weyl-orbit characters
which are given by, for instance,
\ba
\WOaEei{2}{0}{0}{0}{0}{0}{0}{0}{0}(\tau;m_i)\Eqn{=}P(2m_i;2\tau),\\
\WOaEei{2}{1}{0}{0}{0}{0}{0}{0}{0}(\tau;m_i)\Eqn{=}
{1\over 2}\Bigl(P(m_i;\tau/2)+P(m_i;\tau/2+1/2)\Bigr)-P(2m_i;2\tau),\\
\WOaEei{2}{0}{0}{0}{0}{0}{0}{0}{1}(\tau;m_i)\Eqn{=}
{1\over 2}\Bigl(P(m_i;\tau/2)-P(m_i;\tau/2+1/2)\Bigr)
\ea
at level-two. Then by taking the limit $q\to 0$ we find
\ba
a_2(\tau;m_i)\Eqn{=}\frac{1}{24 E_4(\tau)\Delta(\tau)}
\biggl(
 \left(-16{E_4(\tau)}+9{\theta_2(\tau)}^8\right)
 \WOaEei{2}{1}{0}{0}{0}{0}{0}{0}{0}(\tau;m_i)\no\\
&&\hspace{2mm}
+9\left({\theta_3(\tau)}^8-{\theta_4(\tau)}^8\right)
\WOaEei{2}{0}{0}{0}{0}{0}{0}{0}{1}(\tau;m_i)
-135{\theta_2(\tau)}^8\WOaEei{2}{0}{0}{0}{0}{0}{0}{0}{0}(\tau;m_i)
\biggr)\\
\Eqn{\to}
- \tfrac{2}{3}         \WOEei{1}{0}{0}{0}{0}{0}{0}{0}(m_i)%2 - 4
+ 12                   \WOEei{0}{0}{0}{0}{0}{0}{0}{1}(m_i)%2 - 2
-1440
=A_2(m_i),\\[2ex]
b_2(\tau;m_i)\Eqn{=}\frac{1}{576{E_4(\tau)}^2\Delta(\tau)}\no\\
&&\hskip-3mm\times
\biggl(
  \left({\theta_3(\tau)}^4+{\theta_4(\tau)}^4\right)
  \left(-16{E_4(\tau)}^2-15{E_4(\tau)}{\theta_2(\tau)}^8
  +45{\theta_2(\tau)}^{16}\right)
\WOaEei{2}{1}{0}{0}{0}{0}{0}{0}{0}(\tau;m_i)\no\\
&&\hskip3mm
+9{\theta_2(\tau)}^4\left(-12{E_4(\tau)}^2+25{E_4(\tau)}{\theta_2(\tau)}^8
-15{\theta_2(\tau)}^{16}\right)
  \WOaEei{2}{0}{0}{0}{0}{0}{0}{0}{1}(\tau;m_i)\no\\
&&\hskip3mm
+135{\theta_2(\tau)}^8\left({\theta_3(\tau)}^4+{\theta_4(\tau)}^4\right)
  \left(7{E_4(\tau)}-5{\theta_2(\tau)}^8\right)
  \WOaEei{2}{0}{0}{0}{0}{0}{0}{0}{0}(\tau;m_i)
\biggr)\\
\Eqn{\to}
- \tfrac{1}{18}        \WOEei{1}{0}{0}{0}{0}{0}{0}{0}(m_i)%2 - 4
- 3                    \WOEei{0}{0}{0}{0}{0}{0}{0}{1}(m_i)%2 - 2
+ 840
=B_2(m_i).
\ea

Flow of $E_8$ theory to theories with smaller symmetry groups has been 
extensively discussed in the literature \cite{MS,DKV,MNWa,YY}.
In Appendix C we present five-dimensional curves with manifest $E_n$ symmetry 
for $n=6,7$.
 
\subsection{Four-dimensional Curve}

In order to derive four-dimensional curve we first 
have to reinstate the radius $R$ of the
5th dimension and redefine the mass parameter as
\be 
m_i \rightarrow Rm_i, \hskip5mm i=1,2,\ldots ,8.
\ee
Now the new mass parameters carry the dimension of mass. 
Four-dimensional limit is obtained at $R \rightarrow 0$. Thus we expand the
characters $w_{\rho}$ in power series of $R$. Lower order terms of
the series are canceled if we make a suitable shift of the parameter as
\be
u\rightarrow u+6c_2R^2-{1\over 4}{c_2}^2R^4+{1\over 240}{c_2}^3R^6
\ee
where  $c_2$ is the second order Casimir invariant
\be
c_2=\sum_{i=1}^8 m_i^2.
\ee
When we rescale variables as
\be
x\rightarrow R^{10}x,\hskip3mm y \rightarrow R^{15}y, \hskip3mm u\rightarrow 
R^6u,
\ee
all the terms in the curve cancel up to order $R^{29}$ and the $E_8$ curve of
4-dimensional theory \cite{MNb,NTY} appears as the coefficient of $R^{30}$.

Flow of $E_8$ theory to 
theories with lower symmetry has been discussed in \cite{MNa,MNb}.

\subsection{Discussions}

In this paper we have constructed the six-dimensional 
Seiberg--Witten curve
which amounts to an equation describing the complex structure of 
${1\over 2}K_3$\,: by applying
the mirror symmetry technique one can generate the number of holomorphic
curves and gauge theory partition functions on ${1\over2}K_3$ at any higher 
order.
It will be interesting to see if a similar construction of SW curves 
is possible for 
manifolds other than the ${1\over 2}K_3$ surface. 

In \cite{GMS} it was suggested that $SL(2,{\bf Z})$ symmetry of 
$N_f=4$ four-dimensional SW theory should be identified as the $SL(2,{\bf Z})$ 
symmetry of the torus
$T^2$ of six-dimensional theory. Thus there must be a way to
obtain four-dimensional $N_f=4$ theory directly from the six-dimensional
theory without going through 
the degenerate limit of $T^2$ as described in the previous sections.
It is interesting to see if there in fact exists such a direct link between
four- and six-dimensional theories.

\subsection*{Acknowledgements}
K.S. would like to thank M.~Naka, Y.~Ohtake, S.~Terashima
for valuable discussions. Research of T.E. is supported in part by the
fund for the Special Priority Area no.707 ``Supersymmetry and Unified Theory 
of Elementary Particles'' by the Japan Ministry of Education.

\newpage

\renewcommand{\thesubsection}{Appendix \Alph{subsection}.}
\setcounter{subsection}{0}
\setcounter{equation}{0}

\subsection{Coefficient Functions of Six-dimensional SW Curve \label{app6dim}}
\renewcommand{\thesubsection}{\Alph{subsection}.}
\ba
a_4\Eqn{=}
\frac{1}{E_4\Delta^2} \biggl[
 16            \theta_4(2\tau)^8 P(4m_i;4\tau)
+\frac{1}{256} \theta_2\Bigl(\frac{\tau}{2}\Bigr)^8
  P\Bigl(m_i;\frac{\tau}{4}\Bigr)\no\\
&&\hspace{15mm}
+\frac{1}{256} \theta_2\Bigl(\frac{\tau+1}{2}\Bigr)^8
  P\Bigl(m_i;\frac{\tau+1}{4}\Bigr)
+\frac{1}{256} \theta_2\Bigl(\frac{\tau+2}{2}\Bigr)^8
  P\Bigl(m_i;\frac{\tau+2}{4}\Bigr)\no\\
&&\hspace{15mm}
+\frac{1}{256} \theta_2\Bigl(\frac{\tau+3}{2}\Bigr)^8
   P\Bigl(m_i;\frac{\tau+3}{4}\Bigr)
+              \theta_4(2\tau)^8
   P\Bigl(2m_i;\tau+\frac{1}{2}\Bigr)\no\\
&&\hspace{15mm}
+E_4 P(2m_i;\tau)
-\frac{3}{2}\Delta^2{a_2}^2
+\frac{3}{8} E_4\Delta a_2 {b_1}^2
-\frac{9}{128} {E_4}^2{b_1}^4\biggr],\\
\no\\
b_4\Eqn{=}
\frac{1}{E_4\Delta^2}\biggl[
f_{b4,0}(\tau) P(4 m_i;4\tau)
+f_{b4,1}(\tau) P\Bigl(m_i;\frac{\tau}{4}\Bigr)\no\\
&&\hspace{15mm}
+f_{b4,1}(\tau+1) P\Bigl(m_i;\frac{\tau+1}{4}\Bigr)
+f_{b4,1}(\tau+2) P\Bigl(m_i;\frac{\tau+2}{4}\Bigr)\no\\
&&\hspace{15mm}
+f_{b4,1}(\tau+3) P\Bigl(m_i;\frac{\tau+3}{4}\Bigr)
+f_{b4,2}(\tau) P\Bigl(2 m_i;\tau+\frac{1}{2}\Bigr)\no\\
&&\hspace{15mm}
+\frac{5}{48}E_6 P(2 m_i;\tau)
-\frac{1}{96}E_4 E_6\Delta b_3 b_1
-3 E_4\Delta b_2 {b_1}^2
-\frac{55}{384}E_4 E_6 {b_1}^4
\biggr]\\
\mbox{where}
&&f_{b4,0}(\tau)=-\frac{8}{9}\Bigl(32\theta_3(2\tau)^{12}
  -75\theta_3(2\tau)^4\theta_4(2\tau)^8+2\theta_4(2\tau)^{12}\Bigr),\\
&&f_{b4,1}(\tau)=\frac{1}{2^{11}\cdot 9}\Bigl(32\theta_3(\tau/2)^{12}
  -75\theta_3(\tau/2)^4\theta_2(\tau/2)^8+2\theta_2(\tau/2)^{12}\Bigr),\\
&&f_{b4,2}(\tau)=-\frac{1}{18}\Bigl(32\theta_2(2\tau)^{12}
  -75\theta_2(2\tau)^4\theta_4(2\tau)^8-2\theta_4(2\tau)^{12}\Bigr),\\
\no\\
b_5\Eqn{=}
\frac{1}{{E_4}^2\Delta^3}\biggl[
f_{b5,0}(\tau) P(5 m_i;5\tau)
+f_{b5,1}(\tau) P\Bigl(m_i;\frac{\tau}{5}\Bigr)
+f_{b5,1}(\tau+1) P\Bigl(m_i;\frac{\tau+1}{5}\Bigr)\no\\
&&\hspace{20mm}
+f_{b5,1}(\tau+2) P\Bigl(m_i;\frac{\tau+2}{5}\Bigr)
+f_{b5,1}(\tau+3) P\Bigl(m_i;\frac{\tau+3}{5}\Bigr)\no\\
&&\hspace{20mm}
+f_{b5,1}(\tau+4) P\Bigl(m_i;\frac{\tau+4}{5}\Bigr)\no\\
&&\hspace{15mm}
+\biggl(
g_{b5,0}(\tau) P(4 m_i;4\tau)
+g_{b5,1}(\tau) P\Bigl(m_i;\frac{\tau}{4}\Bigr)\no\\
&&\hspace{20mm}
+g_{b5,1}(\tau+1) P\Bigl(m_i;\frac{\tau+1}{4}\Bigr)
+g_{b5,1}(\tau+2) P\Bigl(m_i;\frac{\tau+2}{4}\Bigr)\no\\
&&\hspace{20mm}
+g_{b5,1}(\tau+3) P\Bigl(m_i;\frac{\tau+3}{4}\Bigr)
+g_{b5,2}(\tau) P\Bigl(2 m_i;\tau+\frac{1}{2}\Bigr)\biggr)P(m_i;\tau)\biggr]
\no\\
&&
+\frac{1}{\Delta^3}\biggl[
 \frac{3}{2}\Delta^2 b_3{b_1}^2
+\frac{3}{16}E_6\Delta b_2{b_1}^3
+\frac{1}{9216}(53{E_4}^3+67{E_6}^2){b_1}^5
\biggr]\\
\mbox{where}
&&f_{b5,0}(\tau)=\frac{2}{3}\eta(\tau)^{16}\eta(5\tau)^{16}\no\\
&&\hspace{14mm}\times
\biggl(5^{11}+4\cdot 5^9\frac{\eta(\tau)^{6}}{\eta(5\tau)^{6}}
             +22\cdot 5^6\frac{\eta(\tau)^{12}}{\eta(5\tau)^{12}}
             +4\cdot 5^4\frac{\eta(\tau)^{18}}{\eta(5\tau)^{18}}
             -31        \frac{\eta(\tau)^{24}}{\eta(5\tau)^{24}}\biggr),\\
&&f_{b5,1}(\tau)=\frac{2}{3}\eta(\tau)^{16}\eta(\tau/5)^{16}\no\\
&&\hspace{14mm}\times
\biggl(\frac{1}{5}+4\frac{\eta(\tau)^{6}}{\eta(\tau/5)^{6}}
             +22\frac{\eta(\tau)^{12}}{\eta(\tau/5)^{12}}
             +20\frac{\eta(\tau)^{18}}{\eta(\tau/5)^{18}}
             -31\frac{\eta(\tau)^{24}}{\eta(\tau/5)^{24}}\biggr),\\
&&g_{b5,0}(\tau)=
\frac{2}{3}\theta_3(2\tau)^4\theta_4(2\tau)^8
  \Bigl(-1024\theta_3(2\tau)^{12}
        +1664\theta_3(2\tau)^8\theta_4(2\tau)^4\no\\
&&\hspace{70mm}
        - 644\theta_3(2\tau)^4\theta_4(2\tau)^8
        +  55\theta_4(2\tau)^{12}\Bigr),\\
&&g_{b5,1}(\tau)=
\frac{1}{2^{19}\cdot 3}\theta_3(\tau/2)^4\theta_2(\tau/2)^8
  \Bigl(-1024\theta_3(\tau/2)^{12}
        +1664\theta_3(\tau/2)^8\theta_2(\tau/2)^4\no\\
&&\hspace{65mm}
        - 644\theta_3(\tau/2)^4\theta_2(\tau/2)^8
        +  55\theta_2(\tau/2)^{12}\Bigr),\\
&&g_{b5,2}(\tau)=
-\frac{1}{24}\theta_2(2\tau)^4\theta_4(2\tau)^8
 \Bigl( 1024\theta_2(2\tau)^{12}
       +1664\theta_2(2\tau)^8\theta_4(2\tau)^4\no\\
&&\hspace{70mm}
       + 644\theta_2(2\tau)^4\theta_4(2\tau)^8
       +  55\theta_4(2\tau)^{12}\Bigr),\\
\no\\
b_6\Eqn{=}
\frac{1}{E_4\Delta^3}\biggl[
f_{b6,0}(\tau) P(6 m_i;6\tau)
+f_{b6,1}(\tau) P\Bigl(m_i;\frac{\tau}{6}\Bigr)
+f_{b6,1}(\tau+1) P\Bigl(m_i;\frac{\tau+1}{6}\Bigr)\no\\
&&\hspace{15mm}
+f_{b6,1}(\tau+2) P\Bigl(m_i;\frac{\tau+2}{6}\Bigr)
+f_{b6,1}(\tau+3) P\Bigl(m_i;\frac{\tau+3}{6}\Bigr)\no\\
&&\hspace{15mm}
+f_{b6,1}(\tau+4) P\Bigl(m_i;\frac{\tau+4}{6}\Bigr)
+f_{b6,1}(\tau+5) P\Bigl(m_i;\frac{\tau+5}{6}\Bigr)\no\\
&&\hspace{15mm}
+f_{b6,2}(\tau) P\Bigl(3 m_i;\frac{3\tau}{2}\Bigr)
+f_{b6,2}(\tau+1) P\Bigl(3 m_i;\frac{3\tau+1}{2}\Bigr)
+f_{b6,3}(\tau) P\Bigl(2 m_i;\frac{2\tau}{3}\Bigr)\no\\
&&\hspace{15mm}
+f_{b6,3}(\tau+1) P\Bigl(2 m_i;\frac{2\tau+2}{3}\Bigr)
+f_{b6,3}(\tau+2) P\Bigl(2 m_i;\frac{2\tau+1}{3}\Bigr)\biggr]\no\\
\Eqn{+}\frac{a_2}{E_4\Delta^2}\biggl[
g_{b6,0}(\tau) P(4 m_i;4\tau)
+g_{b6,1}(\tau) P\Bigl(m_i;\frac{\tau}{4}\Bigr)\no\\
&&\hspace{15mm}
+g_{b6,1}(\tau+1) P\Bigl(m_i;\frac{\tau+1}{4}\Bigr)
+g_{b6,1}(\tau+2) P\Bigl(m_i;\frac{\tau+2}{4}\Bigr)\no\\
&&\hspace{15mm}
+g_{b6,1}(\tau+3) P\Bigl(m_i;\frac{\tau+3}{4}\Bigr)
+g_{b6,2}(\tau) P\Bigl(2 m_i;\tau+\frac{1}{2}\Bigr)\biggr]\no\\
\Eqn{+}\frac{1}{\Delta^3}\biggl[
  -\frac{83}{7344} {E_6}{\Delta}^2b_5{b_1}
  +\frac{83}{408} {\Delta}^2b_4{b_1}^2
  +\frac{1}{34} {\Delta}^2{b_3}{b_2}{b_1}
  +\frac{29}{6528} {E_6}{\Delta}{b_3}{b_1}^3
  +\frac{669}{1088} {\Delta}{b_2}{b_1}^4\no\\
&&\hspace{8mm}
  -\frac{5}{7344} {E_4}{\Delta}^2{a_3}{a_2}{b_1}
  -\frac{419}{235008} {E_4}^2{\Delta}{a_3}{b_1}^3
  -\frac{1}{36864} {E_4}{E_6}{\Delta}{a_2}^2{b_1}^2\no\\
&&\hspace{8mm}
  +\frac{1}{1536} {E_4}^2{\Delta}{a_2}{b_2}{b_1}^2
  -\frac{1}{256} {E_6}{\Delta}{b_2}^2{b_1}^2
  +\frac{1215}{69632} {E_6}{b_1}^6\biggr]\\
\mbox{where}
&&f_{b6,0}(\tau)=
-\frac{4}{17}\Bigl(h_0(\tau)+h_0(2\tau)\Bigr)
\Bigl(h_0(\tau)-2 h_0(2\tau)\Bigr)^2\no\\
&&\hspace{14mm}\times
\Bigl(27{h_0(\tau)}^3+84{h_0(\tau)}^2 h_0(2\tau)
+72 h_0(\tau){h_0(2\tau)}^2-32{h_0(2\tau)}^3\Bigr),\\
&&f_{b6,1}(\tau)=
\frac{1}{297432}\Bigl(2 h_0(\tau/3)+h_0(\tau/6)\Bigr)
\Bigl(h_0(\tau/3)-h_0(\tau/6)\Bigr)^2\no\\
&&\hspace{8mm}\times
\Bigl(27{h_0(\tau/3)}^3+42{h_0(\tau/3)}^2 h_0(\tau/6)
+18 h_0(\tau/3){h_0(\tau/6)}^2-4{h_0(\tau/6)}^3\Bigr),\\
&&f_{b6,2}(\tau)=
\frac{1}{136}\Bigl(-2 h_0(\tau)+h_0(\tau/2)\Bigr)
\Bigl(h_0(\tau)+h_0(\tau/2)\Bigr)^2\no\\
&&\hspace{14mm}\times
\Bigl(27{h_0(\tau)}^3-42{h_0(\tau)}^2 h_0(\tau/2)
+18 h_0(\tau){h_0(\tau/2)}^2+4{h_0(\tau/2)}^3\Bigr),\\
&&f_{b6,3}(\tau)=
\frac{4}{37179}\Bigl(h_0(\tau/3)-h_0(2\tau/3)\Bigr)
\Bigl(h_0(\tau/3)+2 h_0(2\tau/3)\Bigr)^2\no\\
&&\hspace{13mm}\times
\Bigl(27{h_0(\tau/3)}^3-84{h_0(\tau/3)}^2 h_0(2\tau/3)
+72 h_0(\tau/3){h_0(2\tau/3)}^2+32{h_0(2\tau/3)}^3\Bigr),\no\\
\\
&&g_{b6,0}(\tau)=
-\frac{640}{51} {\theta_3(2\tau)}^4+\frac{32}{9} {\theta_4(2\tau)}^4,\\
&&g_{b6,1}(\tau)=
\frac{5}{408} {\theta_3(\tau/2)}^4-\frac{1}{288} {\theta_2(\tau/2)}^4,\\
&&g_{b6,2}(\tau)=
-\frac{40}{51} {\theta_2(2\tau)}^4-\frac{2}{9} {\theta_4(2\tau)}^4.
\ea\\

\renewcommand{\thesubsection}{Appendix \Alph{subsection}.}
\subsection{Coefficient Functions of 5d SW Curve
in terms of \\ Characters of Fundamental Representations \label{app5dim}}
\renewcommand{\thesubsection}{\Alph{subsection}.}
\ba
A_0\Eqn{=}\tfrac{1}{12}, \hskip3mm
%A_1=\tchara{0}, \hskip3mm
A_2=-\tfrac{2}{3}{w_1}+\tchara{12}{w_8}-\tchara{1440}, \hskip3mm
A_3=-\tchara{2}{w_2}+\tchara{96}{w_1}-\tchara{1152}{w_8}
  +\tchara{103680},\\
A_4\Eqn{=}\tfrac{4}{3}{w_1}^2-\tchara{4}{w_3}-\tchara{16}{w_6}
  -\tchara{48}{w_1}{w_8}-\tchara{144}{w_8}^2\no\\
&&+\tchara{400}{w_2}+\tchara{1440}{w_7}+\tchara{1728}{w_1}
  +\tchara{41472}{w_8}-\tchara{2073600},\\
B_0\Eqn{=}\tfrac{1}{216}, \hskip3mm
B_1=-\tchara{4}, \hskip3mm
B_2=-\tfrac{1}{18}{w_1}-\tchara{3}{w_8}+\tchara{840},\\
B_3\Eqn{=}-\tfrac{1}{6}{w_2}-\tchara{4}{w_7}-\tchara{8}{w_1}
  +\tchara{528}{w_8}-\tchara{79680},\\
B_4\Eqn{=}\tfrac{2}{9}{w_1}^2-\tfrac{1}{3}{w_3}-\tfrac{16}{3}{w_6}
  -\tchara{24}{w_1}{w_8}-\tchara{120}{w_8}^2\no\\
&&+\tfrac{424}{3}{w_2}+\tchara{1272}{w_7}+\tchara{4608}{w_1}
  -\tchara{25920}{w_8}+\tchara{3939840},\\
B_5\Eqn{=}\tfrac{2}{3}{w_1}{w_2}-\tchara{4}{w_5}-\tchara{16}{w_1}{w_7}
  +\tchara{64}{w_2}{w_8}+\tchara{288}{w_7}{w_8}
  -\tchara{96}{w_1}^2-\tchara{60}{w_3}-\tchara{160}{w_6}
  +\tchara{3456}{w_8}^2\no\\
&&+\tchara{800}{w_2}-\tchara{24480}{w_7}-\tchara{108480}{w_1}
  +\tchara{933120}{w_8}-\tchara{97873920},\\
B_6\Eqn{=}-\tfrac{8}{27}{w_1}^3+{w_2}^2+\tfrac{4}{3}{w_1}{w_3}
  -\tchara{4}{w_4}-\tfrac{32}{3}{w_1}{w_6}-\tchara{48}{w_1}^2{w_8}
  +\tchara{48}{w_2}{w_7}+\tchara{288}{w_7}^2-\tchara{40}{w_3}{w_8}\no\\
&&-\tchara{480}{w_6}{w_8}-\tchara{2592}{w_1}{w_8}^2-\tchara{9792}{w_8}^3
  +\tfrac{1124}{3}{w_1}{w_2}+\tchara{548}{w_5}+\tchara{6688}{w_1}{w_7}
  +\tchara{1884}{w_2}{w_8}\no\\
&&+\tchara{25632}{w_7}{w_8}+\tchara{24576}{w_1}^2+\tchara{12920}{w_3}
  +\tchara{88320}{w_6}+\tchara{578688}{w_1}{w_8}+\tchara{1714176}{w_8}^2\no\\
&&-\tchara{1694400}{w_2}-\tchara{8460000}{w_7}-\tchara{30102720}{w_1}
  -\tchara{104198400}{w_8}+\tchara{721612800}
\ea
where $w_1,\ldots,w_8$ denote $E_8$ Weyl-orbit characters of fundamental
representations and are defined as
\ba
&&
w_1=\WOEei{1}{0}{0}{0}{0}{0}{0}{0}, \hskip3mm
w_2=\WOEei{0}{1}{0}{0}{0}{0}{0}{0}, \hskip3mm
w_3=\WOEei{0}{0}{1}{0}{0}{0}{0}{0}, \hskip3mm
w_4=\WOEei{0}{0}{0}{1}{0}{0}{0}{0},\no\\
&&
w_5=\WOEei{0}{0}{0}{0}{1}{0}{0}{0}, \hskip3mm
w_6=\WOEei{0}{0}{0}{0}{0}{1}{0}{0}, \hskip3mm
w_7=\WOEei{0}{0}{0}{0}{0}{0}{1}{0}, \hskip3mm
w_8=\WOEei{0}{0}{0}{0}{0}{0}{0}{1}.
\ea\\

\renewcommand{\thesubsection}{Appendix \Alph{subsection}.}
\subsection{Five-dimensional $E_n$ SW Curve}
\renewcommand{\thesubsection}{\Alph{subsection}.}
$E_7$ curve:
\ba
y^2\Eqn{=}4 x^3-\Bigl(
\tfrac{1}{12}u^4
+(-\tfrac{2}{3}{w_1}+\tchara{12})u^2
-\tchara{2}{w_2}u
+(\tfrac{4}{3}{w_1}^2-\tchara{4}{w_3}-\tchara{16}{w_6}
  -\tchara{48}{w_1}-\tchara{144})\Bigr)x\no\\
&&-\Bigl(\tfrac{1}{216}u^6
+(-\tfrac{1}{18}{w_1}-\tchara{3})u^4
+(-\tfrac{1}{6}{w_2}-\tchara{4}{w_7})u^3
+(\tfrac{2}{9}{w_1}^2-\tfrac{1}{3}{w_3}-\tfrac{16}{3}{w_6}
  -\tchara{24}{w_1}-\tchara{120})u^2\no\\
&&\hspace{5mm}
  +(\tfrac{2}{3}{w_1}{w_2}-\tchara{4}{w_5}-\tchara{16}{w_1}{w_7}
  +\tchara{64}{w_2}+\tchara{288}{w_7})u
+(-\tfrac{8}{27}{w_1}^3+{w_2}^2+\tfrac{4}{3}{w_1}{w_3}
  -\tchara{4}{w_4}\no\\
&&\hspace{5mm}
  -\tfrac{32}{3}{w_1}{w_6}-\tchara{48}{w_1}^2
  +\tchara{48}{w_2}{w_7}+\tchara{288}{w_7}^2-\tchara{40}{w_3}
  -\tchara{480}{w_6}-\tchara{2592}{w_1}-\tchara{9792})\Bigr)
\ea
where $w_1,\ldots,w_7$ denote $E_7$ Weyl-orbit characters and are defined as
\ba
&&
w_1=\WOEse{1}{0}{0}{0}{0}{0}{0}, \hskip3mm
w_2=\WOEse{0}{1}{0}{0}{0}{0}{0}, \hskip3mm
w_3=\WOEse{0}{0}{1}{0}{0}{0}{0}, \hskip3mm
w_4=\WOEse{0}{0}{0}{1}{0}{0}{0},\no\\
&&
w_5=\WOEse{0}{0}{0}{0}{1}{0}{0}, \hskip3mm
w_6=\WOEse{0}{0}{0}{0}{0}{1}{0}, \hskip3mm
w_7=\WOEse{0}{0}{0}{0}{0}{0}{1}.
\ea
\\
$E_6$ curve:
\ba
y^2=4 x^3\Eqn{-}\Bigl(
\tfrac{1}{12}u^4-\tfrac{2}{3}{w_1}u^2-\tchara{2}{w_2}u
+(\tfrac{4}{3}{w_1}^2-\tchara{4}{w_3}-\tchara{16}{w_6})\Bigr)x\no\\
\Eqn{-}\Bigl(\tfrac{1}{216}u^6-\tfrac{1}{18}{w_1}u^4
  +(-\tfrac{1}{6}{w_2}-\tchara{4})u^3
  +(\tfrac{2}{9}{w_1}^2-\tfrac{1}{3}{w_3}-\tfrac{16}{3}{w_6})u^2\no\\
&&\hspace{2mm}
  +(\tfrac{2}{3}{w_1}{w_2}-\tchara{4}{w_5}-\tchara{16}{w_1})u \no\\
&&\hspace{2mm}
  +(-\tfrac{8}{27}{w_1}^3+{w_2}^2+\tfrac{4}{3}{w_1}{w_3}-\tchara{4}{w_4}
    -\tfrac{32}{3}{w_1}{w_6}+\tchara{48}{w_2}+\tchara{288})\Bigr)
\ea
where $w_1,\ldots,w_6$ denote $E_6$ Weyl-orbit characters and are defined as
\ba
&&
w_1=\WOEsi{1}{0}{0}{0}{0}{0}, \hskip3mm
w_2=\WOEsi{0}{1}{0}{0}{0}{0}, \hskip3mm
w_3=\WOEsi{0}{0}{1}{0}{0}{0}, \hskip3mm
w_4=\WOEsi{0}{0}{0}{1}{0}{0},\no\\
&&
w_5=\WOEsi{0}{0}{0}{0}{1}{0}, \hskip3mm
w_6=\WOEsi{0}{0}{0}{0}{0}{1}.
\ea

\newpage
\renewcommand{\refname}

\end{document}